\documentclass[twocolumn%
 aip,
 jmp,%
 amsmath,amssymb,
 reprint,%
]{revtex4-2}

\usepackage{graphicx}
\usepackage{dcolumn}
\usepackage{bm}
\usepackage[english]{babel}
\usepackage[colorlinks=true, allcolors=blue]{hyperref}
\usepackage{footnote}
\usepackage{svg}
\usepackage{array}
\newcolumntype{R}[1]{>{\raggedleft\let\newline\\\arraybackslash\hspace{0pt}}m{#1}}

\begin{document}
\title{Small Polaron Formation on the Nb-doped SrTiO$_\textbf{3}$(001) Surface}

\author{Florian Ellinger}
\affiliation{University of Vienna, Faculty of Physics, Center for Computational Materials Science, Vienna, Austria }

\author{Muhammad Shafiq}
\affiliation{Department of Physics, Abbottabad University of Science \& Technology, Abbottabad, Pakistan}

\author{Iftikhar Ahmad}
\affiliation{Gomal University, Dera Ismail Khan, Pakistan}

\author{Michele Reticcioli} 
\affiliation{University of Vienna, Faculty of Physics, Center for Computational Materials Science, Vienna, Austria}

\author{Cesare Franchini}
\affiliation{University of Vienna, Faculty of Physics, Center for Computational Materials Science, Vienna, Austria}
\affiliation{Dipartimento di Fisica e Astronomia, Universit\`{a} di Bologna, 40127 Bologna, Italy}

\date{\today}

\begin{abstract}
    The cubic perovsike strontium titanate SrTiO$_3$ (STO) is one of the most studied, polarizable transition metal oxides. When excess charge is introduced to this material, e.g., through doping or atomic defects, STO tends to host polarons: Quasi-particles formed by excess charge carriers coupling with the crystal phonon field. Their presence alters the materials properties, 
    and is a key for many applications. 
    Considering that polarons form preferentially on or near surfaces, we study small polaron formation at the TiO$_2$ termination of the STO(001) surface \textit{via} density functional theory calculations. 
    We model several supercell slabs of Nb-doped and undoped STO(001) surfaces with increasing size, also considering the recently observed as-cleaved TiO$_2$ terminated surface hosting Sr-adatoms. 
    Our findings suggest that small polarons become less stable at low concentrations of Nb-doping, in analogy with polarons localized in the bulk. Further, we inspect the stability of different polaron configurations with respect to Nb- and Sr-impurities, and discuss their spectroscopic properties.
\end{abstract}

\maketitle

\section{Introduction}

    Transition metal oxide perovskites are an incredibly versatile and intriguing class of materials, with electronic properties relevant for a wide range of applications. 
    In particular, SrTiO$_3$ (STO) is one of the most widely investigated perovskite material, as it is a promising candidate for the use in oxide electronics. It exhibits many properties arising on the surface and at the interface with other materials, including the emergence of a two dimensional electron gas (2DEG), that are potentially useful for technological applications in the field of photoemission~\cite{Kan2005}, (photo-) catalysis~\cite{Liu2008, Guan2014, Shoji2016, Ouhbi2021} and superconductivity~\cite{ Reyren2007, Santander-Syro2011, Wang2014, Li2015}.

    The STO(001) surface has attracted a lot of attention in recent experiments.
    Both, the $1\times1$ unreconstructed TiO$_2$- and SrO-terminations, show a weakly polar character and exhibit surface rumpling, together with a small dipole moment~\cite{Goniakowski1996, Sahoo2018, Noguera2000}.
    The weak polarity imposes challenges on sample preparation, leading to inconsistent and partially contradicting data in literature~\cite{Enterkin2010}.
    Just recently, it was possible to obtain an unreconstructed termination of STO(001) surface with $1 \times 1$ symmetry via cleaving~\cite{Sokolovic2021}. The authors were able to introduce a preferential cleavage plane by putting n-type (Nb-)doped crystals under strain. Their scanning probe microscopy measurements on the as-cleaved surface show well defined, large TiO$_2$ and SrO-terminated terraces, 
    with a canonical amount of 14$\pm$2\% Sr-adatoms and -vacancies, 
    respectively~\cite{Sokolovic2019}.
    The spontaneous formation of these defects during the cleaving process was interpreted as a mechanism for compensating the weak polarity.
    These recent experimental findings motivate a re-exploration of the STO surface
    by first principles calculations 
    to account for doping, as well as defects, at variance with earlier theoretical studies~\cite{Eglitis2008}.

    The intrinsic Sr-adatoms and -vacancies influence the electronic properties of the surface by introducing excess charge carriers to the system. 
    Similar to charge carriers introduced by doping (e.g., substituting Ti by Nb), extra charge from the Sr-defects can couple to the crystal phonon field by inducing local distortions in the lattice. As a consequence, these distortions can create a potential well for the charge to localize in a certain region of the crystal, forming a polaron quasi-particle~\cite{Landau1933, Franchini2021}. Polarons have grown to be a major topic of interest when studying semiconducting and insulating materials, especially oxides. Recent studies reveal that these self-trapped charge carriers influence a wide spectrum of physical and chemical properties of the host material~\cite{Spreafico2014, Rettie2016, Zhang2019}. 
    Depending on the crystal, the coupling to lattice distortions and subsequent self-trapping can occur over different length scales: While small polarons are formed by short-range coupling and are usually confined within one unit-cell, large polarons spread over many unit-cell lengths and are formed in the long-range coupling regime.
    As a particularly interesting behavior, STO shows stable electron polarons but can also exhibit delocalized charge in the bulk crystal, as well as a 2DEG on the surface. Some studies suggest that this could hint towards a small-to-large polaron transition, where small polarons, large polarons, and/or delocalized charge can either form under certain conditions, or even coexist (e.g., depending on the charge carrier doping)~\cite{Hao2015,Jeschke_2015}.

    \begin{figure*}[t]
        \includegraphics[width=\textwidth]{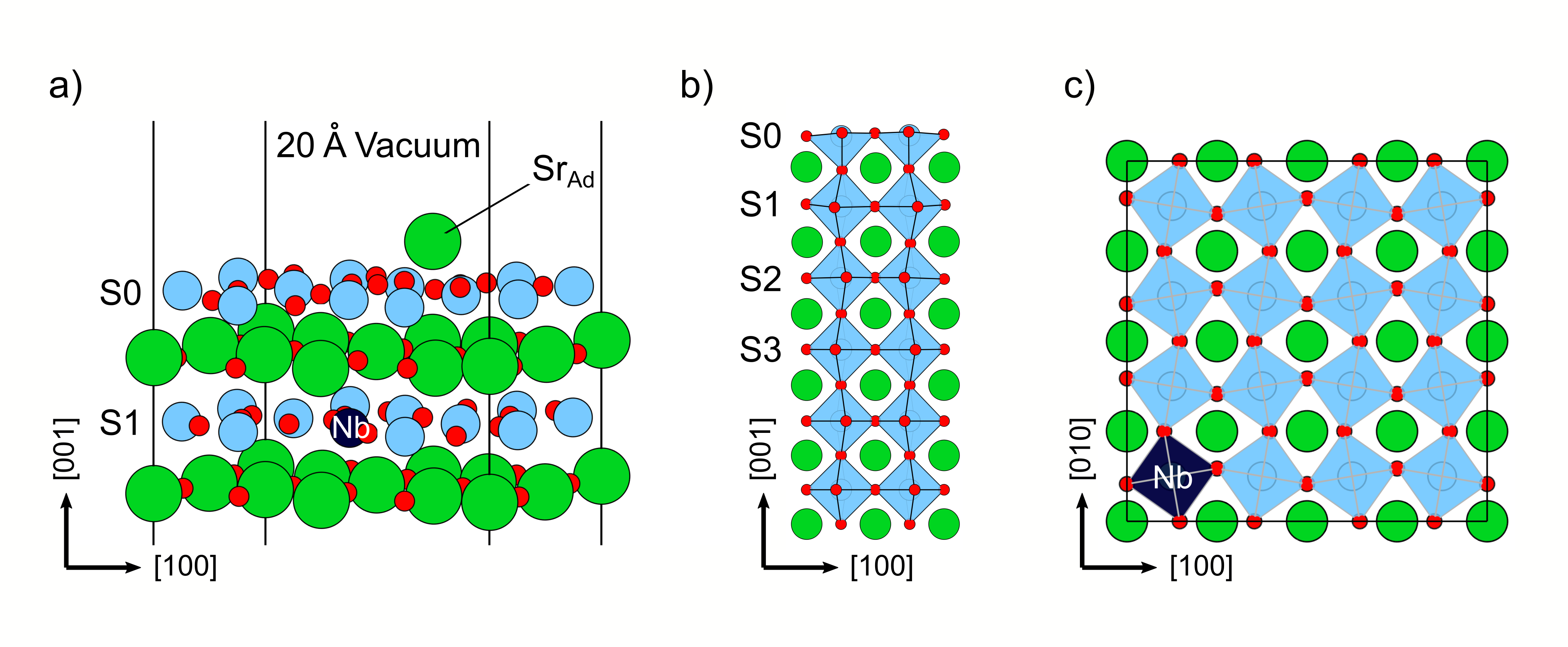}
        \caption{a) Schematic representation of the first two layers of the asymmetric 
        STO(001) surface slab. Nb-doping is placed in the sub-surface (S1) layer and the Sr-adatom at an Sr lattice position on top of the surface. 
        b) Cross-section of a surface slab showing the 6 unit cell layers (6 TiO$_2$ and 6 SrO monolayers, respectively). The O octahedra around Ti sites are rotated in an alternating manner and are slightly tilted due to the anti-ferrodistortive (AFD) lattice. 
        c) Top view of the S1 layer of the $4 \times 4$ supercell. One Ti atom is replaced by Nb for doped systems, O octahedra are rotated in an alternating manner due to the AFD distortions.
        }
        \label{fig:slab}
    \end{figure*}

    We perform a density functional theory (DFT) study of polaronic properties for the TiO$_2$ termination of the STO(001) surface.
    The aim of this study can be divided into two main parts: (1) The investigation of small electron polarons on the surface, introduced in the n-type doped STO crystal by Nb atomic substitution, and (2) the analysis of polaronic properties of the as-cleaved experimental surface including Sr-adatoms, as recently described in Refs.~\cite{Sokolovic2021} and \cite{Sokolovic2019}.
    
    In the first part, a working setup for polaron localization in STO is deducted. Using this setup, polaron formation energies are then calculated for different Nb-doping concentrations. 
    Afterwards, results from the first part will be used for modelling and investigating the as-cleaved TiO$_2$ termination, including Sr-adatoms. We compare different polaron configurations, as well as generate simulated scanning tunnelling microscopy images for different environments. 
    
\section{Methods}

    \subsection{Computational Setup}
    
    Calculations were done using the Vienna \textit{Ab initio} Simulation Package (VASP)~\cite{vasp1, vasp2, vasp3}.
    Projector-Augmented-Wave Perdew-Burke-Ernzerhof potentials (PAW-PBE) were used~\cite{Blochl1994, Perdew1996, Kresse1999}. Since charge localization is spuriously disfavored in standard DFT, we employed an on-site energy correction term within the DFT+$U$ framework to describe small polaron localization. For this, we chose the rotational invariant implementation as introduced by Dudarev \textit{et al.}~\cite{Dudarev1998} with a value of $U - J = 4.5$\,eV, enacted on the $d$-orbitals of Ti and Nb atoms. This is consistent with the choice of $U$ in earlier works using values calculated by constrained random phase approximation (cRPA)~\cite{Hou2010, Choi2013, Hao2015}.

    \subsection{Supercells}

    To model both, the bulk-like and the as-cleaved TiO$_2$ termination of the STO(001) surface, we used asymmetric slabs  with a thickness of 6 unit-cell layers.
    Each unit-cell layer consists of one TiO$_2$- and one SrO-monolayer, respectively, amounting to 12 stacked monolayers in total.
    We adopted the low-temperature, anti-ferrodistortive (AFD) phase of STO, consisting of 20 atoms in the bulk primitive cell,  together
    with experimental lattice parameters ($a = b = 5.5202$\,\r{A}, $c = 7.8067$\,\r{A},~\cite{Yamanaka2002}). 
    The bottom two unit-cell layers were fixed at the AFD bulk structure, while the atom positions in the top four layers were relaxed until the norms of all forces were below 0.01\,eV/\r{A}. For the ionic relaxation a plane-wave energy cut-off of 550\,eV was used, together with a $5\times 5 \times 1$ k-point grid.  The energy cut-off was increased to 700\,eV for the final electronic ground state calculations. 
    Due to STO(001) being a weakly polar surface, a dipole correction along the z-axis (normal to the surface) was applied~\cite{Makov1995, Neugebauer1992}.
    
    Supercells of increasing lateral size were constructed to model the bulk-like $1\times1$ TiO$_2$ termination. For this, the lateral dimensions of the surface slab were scaled by a factor of $\mathrm{N}_\mathrm{unitcell} = \{2, 2\sqrt{2}, 3\sqrt{2}, 4, 4\sqrt{2}\}$.
    This results in cells ranging from 240 atoms for the smallest ($2 \times 2 \times 6$ slab) to 960 atoms for the largest one ($4\sqrt{2} \times 4\sqrt{2} \times 6$ slab).
    
    The Nb-doping was included by substituting one sub-surface (S1) Ti atom, leading to one excess electron in our supercells. The Nb-substitution on different layers (in the absence of charge localization) is less stable, as shown in Tab.~\ref{tab:nb_placement} for the $2\sqrt{2}$ slab.

    The as-cleaved surface was modeled by adding one Sr-adatom on both, a pristine and Nb-doped $2\sqrt{2} \times 2\sqrt{2} \times 6$ slab, corresponding to an adatom coverage of 12.5\%, closely resembling the experimentally observed coverage of 14$\pm$2\%~\cite{Sokolovic2019}.
    For the Nb-doped slab, the Sr-adatom was placed at the maximally allowed distance between the two defects, as this configuration was found to be the energetically most stable (see Supplementary Information for details). A schematic view of the adopted slabs is shown in Fig.~\ref{fig:slab}.

   \begin{table}[h!]
        \caption{Difference in ground state energy for Nb-doping being placed in different layers of the slab. Calculations were done for the slab with lateral dimensions of $2\sqrt{2}$ unit-cell lengths and the excess electron being delocalized.}
        \begin{tabular}{r|R{1cm}R{1cm}R{1cm}R{1cm}}
             Layer & S0 & S1 & S2 & S3 \\
             $\Delta$E (meV)  & +3 & 0 & +61 & +63
             \end{tabular}
        \label{tab:nb_placement}
    \end{table}
    
 \subsection{Polaron Localization}    
    
    The Nb-doping and Sr-adatoms introduce one and two excess charge carriers, respectively, that can localize in the form of polarons.
    The polaron stability can be inspected by calculating the polaron formation energy $E_{pol}$, defined as the difference between the total energy of the system containing localized polarons ($E^{tot}_{loc}$) and the system with the excess charge being delocalized ($E^{tot}_{deloc}$):

    \begin{equation}
        E_{pol} = E^{tot}_{loc} - E^{tot}_{deloc} ~.
    \end{equation}
    
    Different localization sites and polaronic arrangements were inspected. Two different approaches were adopted to induce localization of the small electron polarons at the desired Ti site: (1) A step-wise procedure, where distortions around the localization site are induced by intermediate replacement of the Ti atom with V. After re-substitution, the polaron 
    forms at the selected site.~\cite{Reticcioli2018} (2) Charge localization using an Occupation Matrix Control tool (Occmatrix), where orbital occupations are initialized
    directly through a hard constrain in a preliminary calculation. This way, polaronic charge can be placed in the desired Ti orbital explicitly, and distortions are induced by the excess charge directly~\cite{Allen2014}.
        
    The upsides of the Occmatrix approach are a faster and more straightforward protocol, as well as that it allows for a precise control of the selected trapping site. A downside is, that it needs an initial educated guess for the polaron orbital. These orbital configurations can be extracted from previous calculations with an already localized polaron. Since the characteristic octahedral crystal structure is similar for all calculations done here, the preferred orbitals for polaron localization, governed by the crystal field, do not change drastically between supercells. Further, even starting from a rough initial guess the excess charge typically has enough freedom to find the ground state solution during the final, unconstrained self-consistent calculation, and relaxes into the lowest-energy orbital. This was generally true during our work and was also confirmed by the Occmatrix developers~\cite{Allen2014}. Only in some scenarios, where two solutions are close in energy, the initial orbital guess can lead to different polaron configurations. One critical aspect, which is worth noting, however, is that the induced distortions need to be large enough for the polaron to stay localized. In some calculations it was required to increase the occupation of the polaron orbital in the initial guess slightly, i.e., placing more than one additional electron,
    to guarantee persistent electron trapping.

    Simulated scanning tunnelling microscopy images (simSTM) of the STO(001) surface were obtained by considering the partial charge density of the in-gap polaron states at approximately 3\,\r{A}\ above the surface.

     \begin{figure}[t!]
        \includegraphics[width=0.49\textwidth]{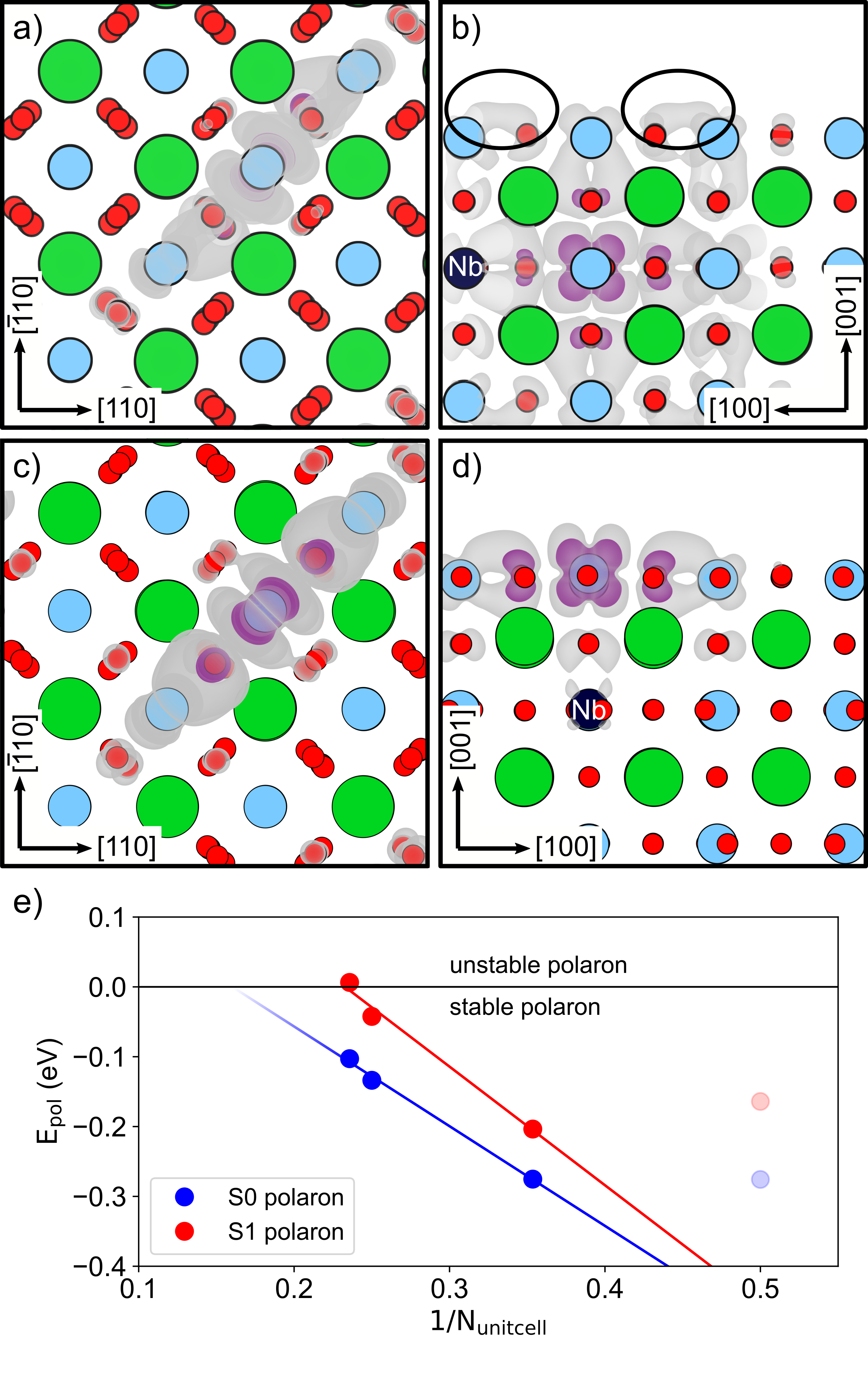}
        \caption{Charge isosurface of single polarons and corresponding formation energies ($\mathrm{E_{pol}}$) for different supercell sizes. a-b) Top- and side-view of the S1 polaron, localized at a Ti site next to Nb. Polaronic charge localized at Ti-O bonds (encircled) creates the strongest signal in  simSTM images discussed in Sec. \ref{sec:simstm}. c-d) Top- and side-view of a S0 polaron, localized at a Ti site on the surface, directly above Nb. e) Polaron formation energies as a function of Nb-doping concentration for S0 and S1 polarons. 
        Surface polarons (S0) are more stable than polarons localized in the S1 layer, where for the latter charge delocalization is favored for the $3\sqrt{2} \times 3\sqrt{2} \times 6$ supercell already. The $2 \times 2$ supercells (light shaded points) are excluded from the linear trend, as for these systems the DOS and isosurfaces suggest polaron self-interaction due to finite size effects.}
        \label{fig:1pol_epol}
    \end{figure}

    \begin{figure}[b!]
        \includegraphics[width=0.49\textwidth]{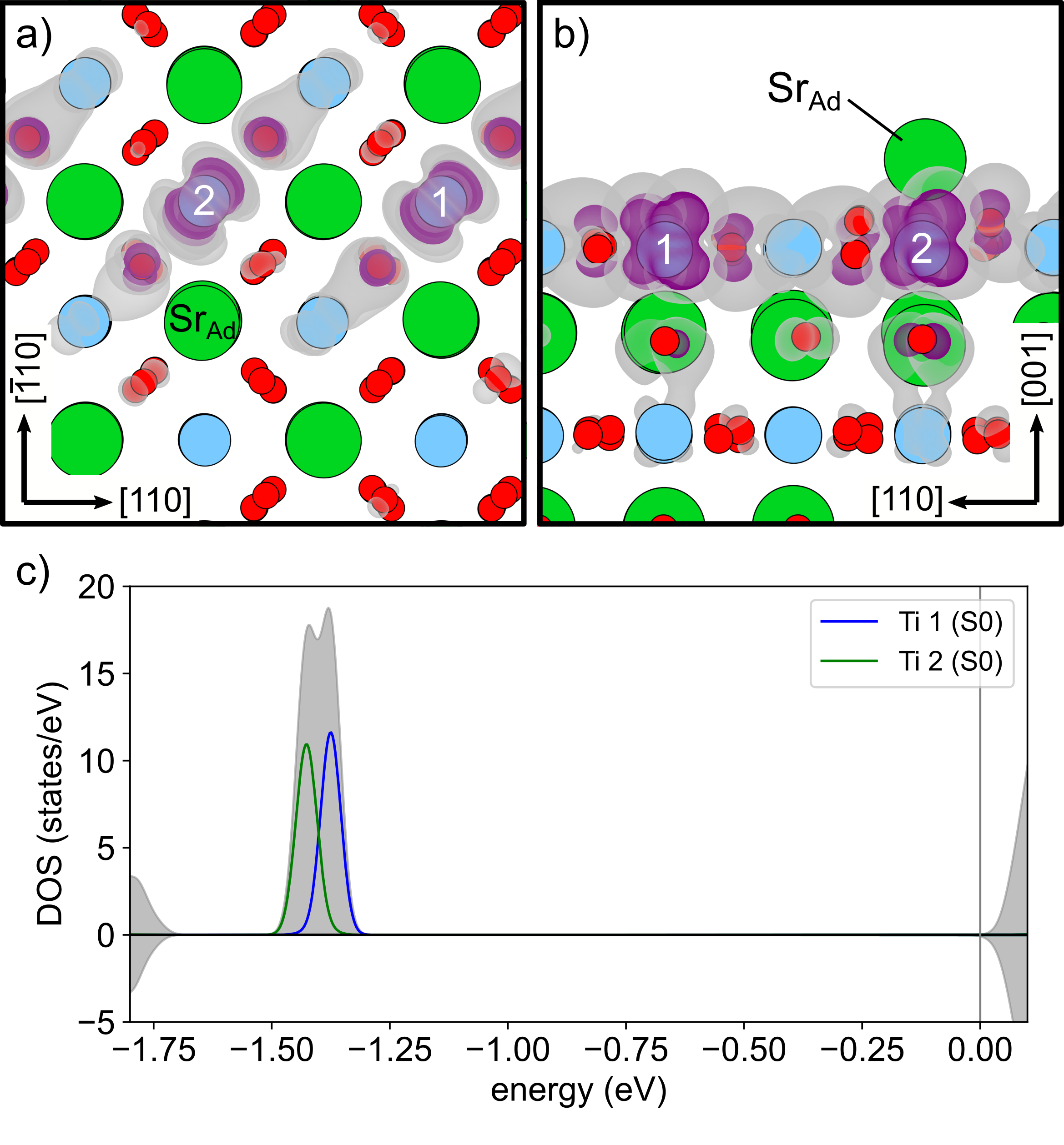}
        \caption{a-b) Polaron charge isosurfaces of the undoped surface slab with Sr-adatom (Sr$_{Ad}$) in top- and side-view, respectively. 
        c) Corresponding in-gap region of the partial density of states (pDOS) of the two polaronic Ti sites (colored full lines), as well as the total DOS (gray shaded area).}
        \label{fig:und_parchg}
    \end{figure}
    
\section{Results}
    
    \subsection{Polaron stability on the bulklike-terminated STO(001)}\label{sec:pol_stab}
    
    We calculated the polaron stability depending on the doping concentration by localizing one polaron in the Nb-doped slabs with varying lateral size (see Fig.~\ref{fig:1pol_epol}).
    The polaron was localized at a Ti site next to (S1 polaron) or above (S0 polaron) the subsurface Nb dopant. These represent the most favorable localization sites as obtained in a preliminary search. For this we have tested all possible Ti sites within the S0 and S1 layers of the $4 \times4 \times 6$ slab (see Supplementary Information for details). Fig.~\ref{fig:1pol_epol}a-d show the polaronic charge density for these two kinds of polaron. In both cases, the polarons occupy d-orbitals of Ti atoms in the t$_{2g}$ symmetry group ($d_{xz}$ and $d_{yz}$ orbitals, preferably). This leads to orbital lobes being aligned with the O octahedron surrounding the polaronic Ti ion (see Fig.~\ref{fig:1pol_epol}a-d).
    We note that it was energetically unfavorable to obtain different solutions, even when making use of the occupation matrix constrains. However, the presence of defects might alter this scenario, as discussed later in Sec. \ref{sec:exp_surf} for the as-cleaved surface with Sr-adatoms.

    The formation energy ($E_{pol}$) trend for S0 and S1 polarons  as a function of Nb-doping is shown in Fig.~\ref{fig:1pol_epol}e.
    Clearly, at high doping level, the stability of both, the S0 and S1 polarons, linearly increases for increasing doping concentration within the investigated regime.
    Conversely, a polaron instability arises at low doping concentrations.  
    The crossover to the unstable regime is reached for the S1 polaron in the $3\sqrt{2} \times 3\sqrt{2} \times 6$ STO(001) supercell, corresponding to a doping concentration of 0.93~\% (1 Nb and 107 Ti ions), while the crossover for the S0 polaron is estimated to take place at a concentration of 0.33~\%. 
    At lower level concentration, polaron formation becomes highly unfavorable: For example, the S1 polaron in the larger $4\sqrt{2} \times 4\sqrt{2} \times 6$ surface slab (doping level of 0.52~\%) shows a positive formation energy of $E_{pol} = 2.346$\,eV (not shown in Fig.~\ref{fig:1pol_epol}).   
    These critical values are in very good agreement with earlier studies investigating the bulk system, where a transition to the unstable regime was found at a doping concentration of 0.8~\%, and attributed to a reduced flexibility of the lattice at lower doping concentrations~\cite{Hao2015}.

    \subsection{The as-cleaved TiO$_2$ termination}\label{sec:exp_surf}

    We modeled the as-cleaved TiO$_2$ termination considering two different setups: (1) An undoped $2\sqrt{2} \times 2\sqrt{2} \times 6$ slab with one Sr-adatom adsorbed, leading to two small electron polarons in the supercell, and (2) an Nb-doped slab of same size together with the adsorbed Sr-adatom, resulting in three excess charge carriers total.
    
    \begin{figure*}[t!]
        \includegraphics[width=\textwidth]{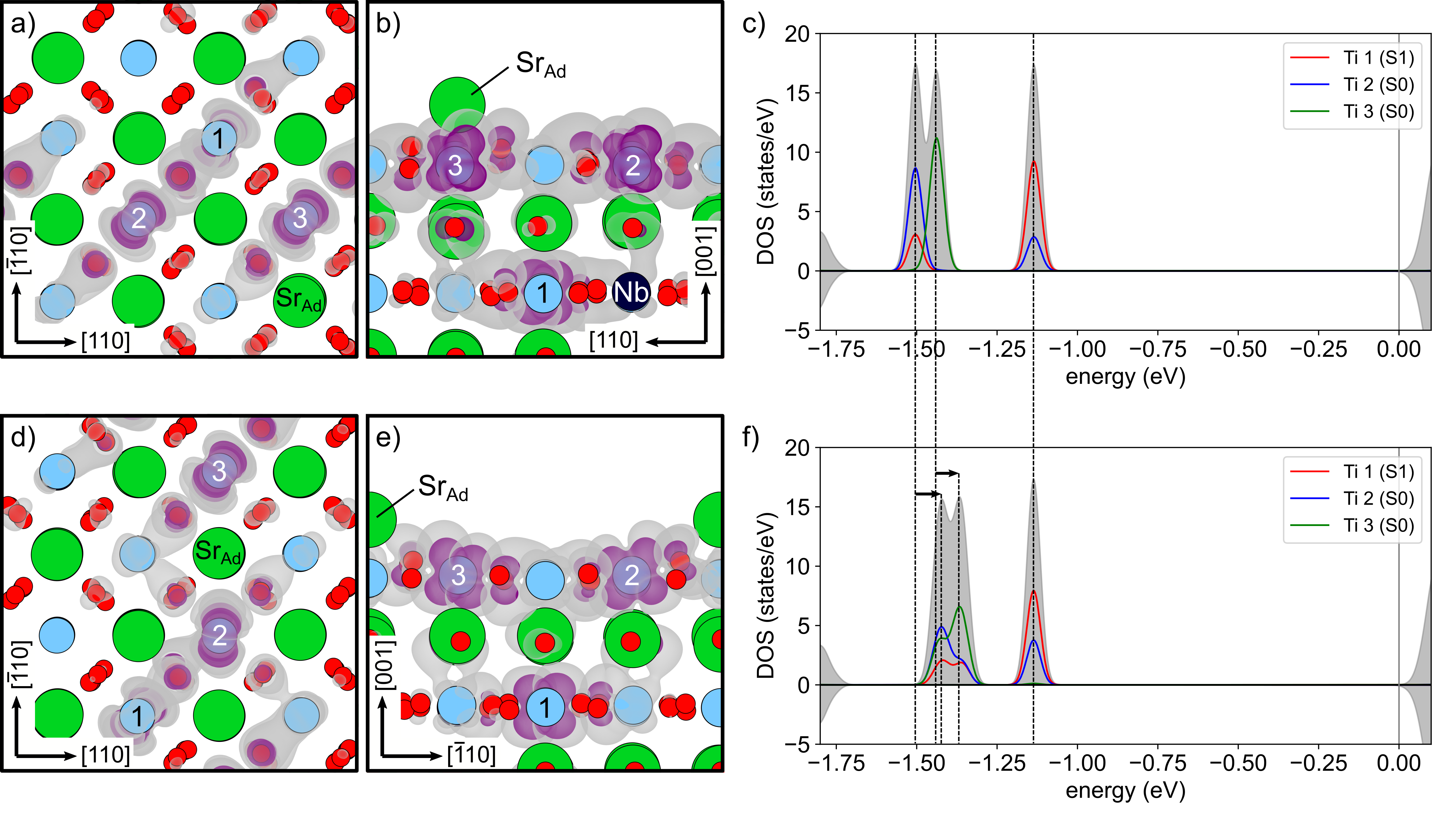}
        \caption{a-b) Top- and side-view of the ground state polaron configuration for the experimental slab. Two polarons are localized in the S0 layer, the third one in the S1 layer.
        c) pDOS plot for the ground state polaron configuration; surface polarons (Ti 2 and 3) appear as distinct peaks and are lower in energy than the sub-surface polaron peak (Ti 1).
        d-e) Top- and side-view for a different three-polaron configuration. This configuration is higher in energy, due to one surface polaron (Ti 2) now being localized further away from Nb, as well as unfavorable polaron orbital rotations. 
        f) Corresponding pDOS, where polaron peaks of the surface polarons (Ti 2 and 3) are shifted higher up in energy compared to the ground state configuration. The pDOS shows a strong charge sharing between polaronic sites, resulting in the formation of a multi-polaron complex.
        }
        \label{fig:exp_surface}
    \end{figure*}
    
        \begin{figure*}[t]
        \centering
        \includegraphics[width=\textwidth]{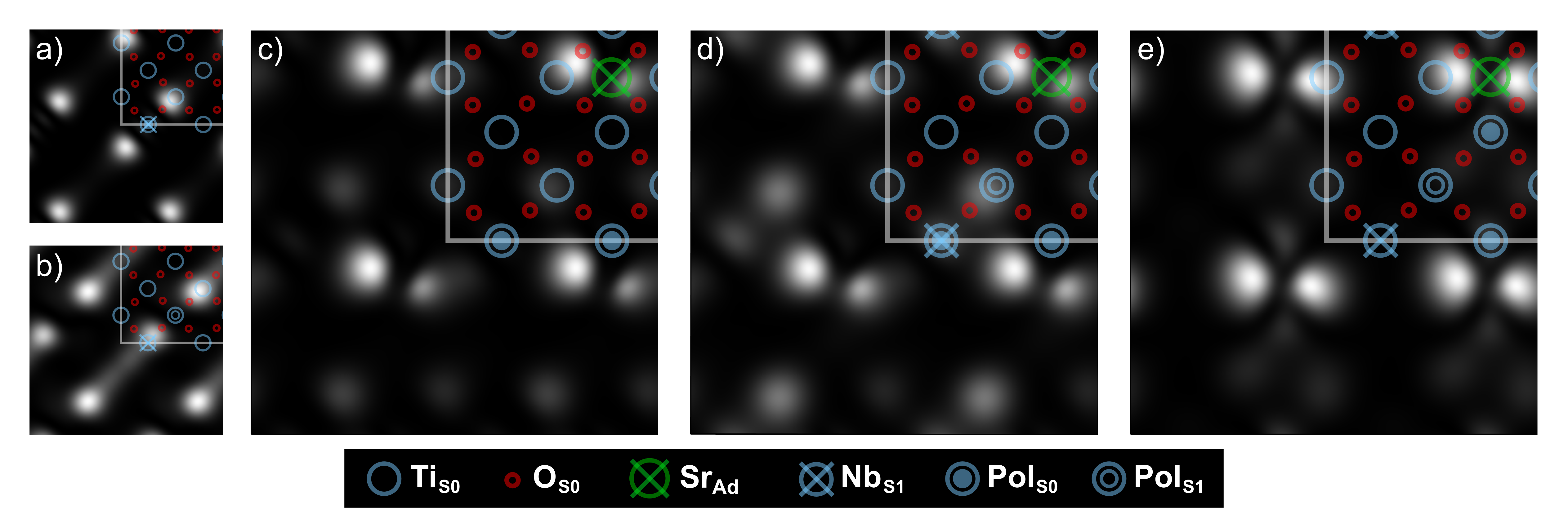}
        \caption{Simulated STM images taken 3\,\r{A}\ above the surface of the $2\sqrt{2} \times 2\sqrt{2} \times 6$ cell. The image is replicated in x- and y-direction for better visualization of the features. A single supercell is indicated by the overlay, in which different defects and polaron positions are marked.
        a) System with one polaron localized in the S0 layer, directly on top of the Nb doping. 
        b) One polaron localized at a Ti atom next the Nb doping  within the S1 layer. 
        c) Two polarons localized in the S0 layer, introduced by the Sr-adatom. 
        d) Ground state configuration for three polarons with both defects present in the supercell, Nb doping and Sr-adatom. One polaron is localized in the S1 layer next to the Nb atom, and two in the S0 layer. 
        e) Another configuration with three polarons for which charge sharing between different polaronic sites is especially pronounced (See Fig. \ref{fig:exp_surface}f). 
        }
        \label{fig:sim_stm}
    \end{figure*}

    For the first case, only the Sr-adatom acts as electron donor and the ground state configuration is represented by two small polarons localizing at S0 Ti atoms (see Fig.~\ref{fig:und_parchg}). This appears as a reasonable solution, considering that S0 polarons are more stable than S1 polarons, as shown in Fig.~\ref{fig:1pol_epol}e for the bulk-like termination. Further, with only two polarons occupying the same layer, there is enough space to prevent strong repulsion between them. The charge of the two polarons is aligned in parallel directions, as shown in Fig.~\ref{fig:und_parchg}a-b: This orientation shows minimal overlap of the polaronic charge densities, minimizing the repulsive interaction.

    The polaron localized at the Ti atom next to the Sr-adatom shows a slightly better formation energy, as indicated by the two in-gap polaron peaks in the density of states in Fig.~\ref{fig:und_parchg}c. The expected pattern would be localizing both (negatively charged) polarons near the (positively charged) Sr-adatom, at Ti atoms on opposite sides of the adatom. However, this configuration is disfavored by 38\,meV, due to polaronic charge of one polaron spreading towards the adatom, creating a repulsive environment for the other (as evident also from the simulated STM images reported in Sec.~\ref{sec:simstm}).
    All other available S0 Ti sites seem to show very similar formation energy for the localization of the second polaron.

    In the Nb-doped slab, the scenario becomes more complicated due to the higher number of charge carriers (one induced by the Nb, two by the Sr-adatom). 
    As discussed for bulk-like termination, the Nb atom creates an energy minimum for polaron localization at its nearest-neighbor (NN) Ti atoms in both, the S1 and S0 layers. 
    The ground state configuration for the Nb doped slab with Sr-adatom is retrieved for two polarons localizing in the S0 layer, one directly on top of the Nb atom and the other at a Ti site next to the Sr-adatom, and one polaron localizing at a Ti site next to Nb in the S1 layer (see Fig. \ref{fig:exp_surface}a-b). 
    The corresponding density of states is shown in Fig. \ref{fig:exp_surface}c. The S1 polaron peak is clearly higher in energy, in line with a less favorable $E_{pol}$ energy discussed for the bulk-like termination. The degeneracy in energy for the two S0 polarons is lifted due to the defects, where the lower-energy peak is associated with the S0 polaron being directly above Nb. 
    Due to the large density of carriers, the polaronic states can get affected by a sizable overlap of the charge densities: The DOS clearly shows that the S1 polaron spreads over the Ti site hosting one of the two S0 polarons (see the red and blue peaks in the DOS).

    For comparison, Fig.~\ref{fig:exp_surface}d-e show a different polaron configuration, characterized by less stable total formation energy, $\Delta E_{pol} = 124$\,meV. The corresponding in-gap DOS (Fig. \ref{fig:exp_surface}f) reveals a similar electronic structure, with the polaronic peaks shifted towards higher energy, in line with the less favorable $E_{pol}$.
    This might arise from an increased polaron-polaron repulsion, due to the enhanced overlap of the polaronic orbitals (see the smeared peaks in the DOS), as well as a less favorable rotation of polaron orbitals.

    \subsection{Simulated STM} \label{sec:simstm}
    
    The simSTM pictures of all previously discussed polaron systems are collected in Fig.~\ref{fig:sim_stm}. Single-polaron systems with one polaron in S0 or S1 are shown in panel a and b, respectively. These are the systems discussed in Sec.~\ref{sec:pol_stab}, using a Nb-doped slab with bulk-like termination. For both, S0 and S1 polarons, a two-dotted signal is observed for each single polaron. The strongest signal originates from charge localized at Ti-O bonds at the surface, however not directly adjacent to the localization site. The bond lobes creating this signal are marked in Fig.~\ref{fig:1pol_epol}b.
    
    Fig~\ref{fig:sim_stm}c depicts the simSTM image of the undoped slab, with only the Sr-adatom present as electron donor. Both polarons are localized in S0, as shown in Fig.~\ref{fig:und_parchg}, and again the strongest signal is located at surface Ti-O bonds. Furthermore, charge is accumulated around the Sr-adatom.

    The ground state configuration for the as-cleaved slab with Sr-adatom and Nb-doping is shown in Fig. \ref{fig:sim_stm}d. This surface slab recreates the experimental surface most accurately of all systems investigated here. The simSTM picture looks very similar to the undoped slab in panel c, with S0 polaron charge accumulating around the Sr-adatom. However, there is also a stronger signal visible at a Ti-O bond further away from the adatom, caused by the additional S1 polaron.
    
    A simSTM picture of the alternative polaronic configuration depicted in Fig.~\ref{fig:exp_surface}d-e is shown in Fig.~\ref{fig:sim_stm}e. This system has enhanced charge accumulation around the Sr-adatom, resulting in a stronger and more symmetric feature in the simSTM signal. The S1 polaron signal is suppressed and the charge most likely incorporated in the strong signal around the adatom. This shows that multiple polarons close to each other may lead to very different patterns in simSTM, where charge spreading between polaronic sites can suppress or enhance certain features in the picture.

\section{Conclusions}

    In the presented work we modeled polaron formation on an unreconstructed STO(001) surface with TiO$_2$ termination, while taking into account chemical doping and defects. Our thorough classification of polarons hosted in this system leads to several relevant findings, all of which could help to unravel the specific roles of these quasi-particles in STO in future theoretical and experimental research.
    
    First, we were able to extend a known trend for STO bulk systems to the STO(001) surface, that polarons tend to be more stable in doping-rich environments. Consequently, when moving towards the dilute limit, e.g., low doping concentration, polarons are progressively destabilized until they become unstable for very sparse doping environments. For polarons localized in the sub-surface (S1) TiO$_2$ layer this cross-over occurs at roughly the same doping concentration as for bulk systems~\cite{Hao2015}, while polarons at the surface (S0) are more stable. This result provides further support to the claimed competition between small and large polaron in this system, or the possible coexistence of localized and delocalized charge in different regions of the crystal. 
    
    Next, we have modelled the recently obtained, as-cleaved STO(001) surface with TiO$_2$ termination, a prime example of a flat (001) perovskite oxide surface.  This was achieved by considering Sr-adatoms and Nb-dopants simultaneously in the same simulated slab.
    Using realistic defect and doping concentrations, our data suggest that under these conditions the STO system hosts three polarons per supercell. The resulting polaron patterns are rationalized by several overlapping effects: On the one hand, the negatively charged small electron polarons energetically favor a separation between each other due to repulsive polaron-polaron interactions, whereas on the other hand they are attracted to the positively charged defects, in this case the positively charged  Nb and Sr ions.
    The resulting configuration is formed by two polarons localized in the S0 layer, and the third polaron being pushed down to the S1 layer.
    Further, polarons localized directly on the surface in the S0 layer are more stable and energetically preferred, suggesting that they can more easily interact with external adsorbates. 
    
    An additional mechanism that influences the stability of the studied multi-polaron system is polaron-charge spreading over several different polaronic Ti sites, as well as polaron-orbital ordering, which lead to rotations and re-orientations of polarons lobes with respect to each other.~\cite{PhysRevB.98.045306} 
    In a very hybridized configuration, one polaron orbital rotates out of the typical ground state position, no longer aligning its orbital lobes with the surrounding TiO$_6$ octahedra. 
    Charge sharing or polaron clustering may be a way to increase polaron stability in a densely populated environment.
    
    The spectroscopic signal of polarons were analyzed by means of simulated STM images. A single polaron, localized either in the S0 or S1 layer, creates a distinct, two-dotted signal, while more complex patterns emerge in environments hosting multiple polarons. Polaronic charge is found primarily in Ti-O bonds at the surface and accumulated around the Sr-adatom, when present. This charge accumulation could lead to altering the adatom appearance in experimental images. Charge spreading between polaronic sites can lead to the suppression or enhancement of certain features in the STM picture, further complicating the interpretation.
    
    Overall our study provides additional theoretical evidence to the polaronic behavior of STO(001) and helps interpreting the growing amount of experimental data measured on its pristine and unreconstructed terminations. Future work will follow this first state analysis, focusing on the interaction between polarons and adsorbates and the possible coexistence of electron and hole small polarons, aspects key to design energy applications based on STO. 
    
\section{Acknowledgements}
    Useful discussions with M. Setvin, U. Diebold, I. Sokolovic and M. Schimd are gratefully acknowledged. 
    
    This work was funded by the FWF project P 32148-N36 SUPER (\textit{Surface science of bulk-terminated cubic perovskite oxides}), and supported by the Austrian BMBWF mobility project CZ15/2021.
    
    MS, IA and CF acknowledge financial support from Higher Education Commission of Pakistan (HEC) project No. 1-8/HEC/HRD/2015/3726.
    The computational results presented have been achieved using the Vienna Scientific Cluster (VSC).
    
\section{References}
    
    \bibliography{ref}

\end{document}


\maketitle

\section*{Polaron Stability: Nb-Pol Distance}

\begin{figure}[h]
\centering
\includegraphics[width=\textwidth]{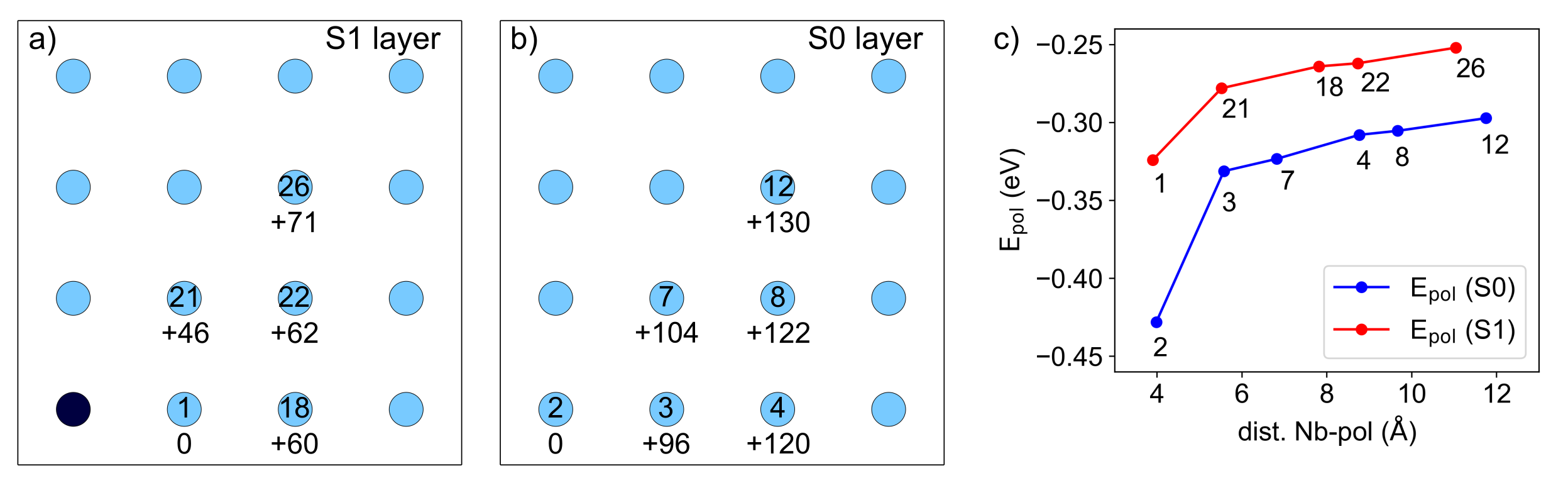}
\caption{\label{fig:doping_distance}a-b) Polaron formation energy for one S1 or one S0 polaron placed at different Ti sites, repectively. Difference in polaron formation energy (E$_{\mathrm{pol}}$) to the ground state placement is given in meV for each inequivalent Ti atom. c) Formation energy in dependence of the distance to the Nb atom, the most stable polarons are found at nearest-neighbor Ti sites of Nb.}
\end{figure}

\newpage

\section*{Placement of Sr Adatom}

\begin{figure}[h]
\centering
\includegraphics[]{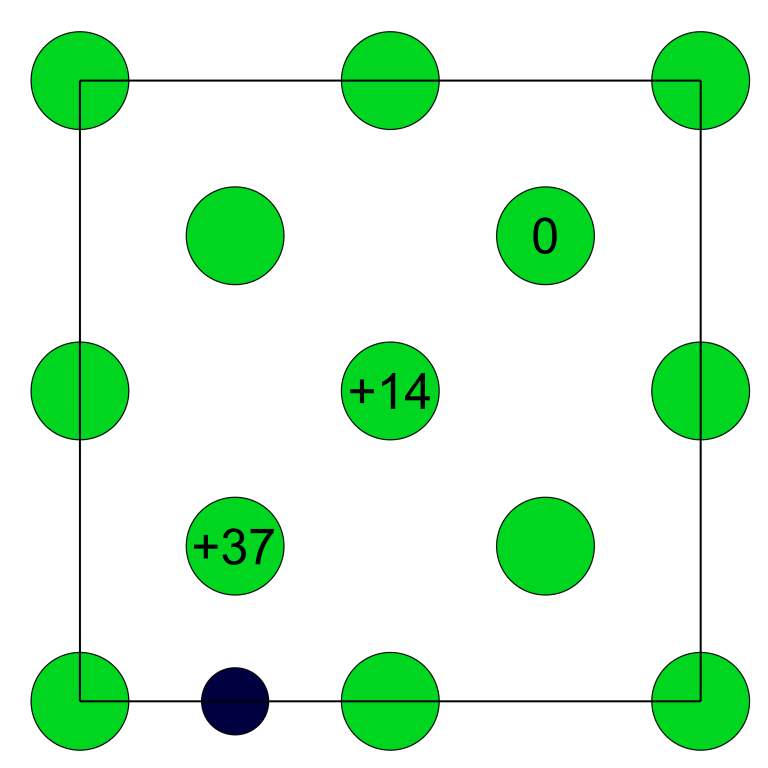}
\caption{\label{fig:sr_placement}Difference in total energy of inequivalent Sr-adatom placements. The Nb-doping atom placed in the S1 layer is shown in dark-blue. The energy difference relative to the ground state placement (at maximum adatom-Nb distance) is given in meV.}
\end{figure}